\documentclass[journal]{IEEEtran}
\usepackage{graphicx}

\ifCLASSINFOpdf
\else
\fi
\begin{document}
%
\title{Experimental wavelength-space division multiplexing of quantum key distribution with classical optical communication over multicore fiber}
%
%
%

\author{Chun~Cai,
        Yongmei~Sun,
        Yongrui~Zhang,
        Peng~Zhang,
        Jianing~Niu,
        and~Yuefeng~Ji
\thanks{Yongmei Sun is with the The State Key Laboratory of Information Photonics and Optical Communications, Beijing University of Posts and Telecommunications, Haidian District, Beijing, 100876, China
e-mail: ymsun@bupt.edu.cn.}
\thanks{Chun Cai, Yongrui Zhang, Peng Zhang, Jianing Niu, and Yuefeng Ji are with Beijing University of Posts and Telecommunications.}
}

\maketitle

\begin{abstract}
We demonstrate quantum key distribution (QKD) with classical signals in a seven-core fiber using dense wavelength division multiplexing. Quantum signals are transmitted in an outer core separately and intercore crosstalk (IC-XT) is the main impairment of them. In order to alleviate IC-XT, we propose a quantum-classical interleave scheme. Then the properties of IC-XT are analyzed based on the measurement results which indicate counter-propagation is a better co-existence method than co-propagation. Finally, we perform QKD experiments in the presence of two classical channels with a channel spacing of 100 GHz between quantum channel and the nearest classical channels. The experiment results prove counter-propagation almost immune to IC-XT, which is consistent with our analysis. Also, the feasibility of the transmission over the range of metropolitan area networks is validated with our scheme.
\end{abstract}


%
\IEEEpeerreviewmaketitle

\section{Introduction}
%
%
%
%
\IEEEPARstart{Q}{uantum} key distribution (QKD) is the most successful application in quantum information science at present. It allows remote parties to establish encryption keys by the laws of physics \cite{Bennet1984Quantum,Gisin2001Quantum}. QKD can detect any eavesdropping attacks and therefore enables information-theoretic communication security \cite{Shor2000Simple}.

There has been a tremendous interest in developing this quantum technology for real-life applications.  In some experiments, QKD was realized using separate fibers rather than by coexisting with the classical channels.
However, due to the high cost of laying fiber resources and the difficulty of synchronization in separate transimission, there is an increasing demand for multiplexing of QKD with traditional optical communications in a single fiber. Wavelength division multiplexing (WDM) is one of the most mature technologies to realize it. The first WDM scheme for QKD and classical channels was proposed by Townsend in 1997 \cite{575910}. Subsequently, quantum channel was set in O-band to reduce impairment from the classical channels which are usually located at C-band \cite{nweke2005experimental,chapuran2009optical,choi2011quantum}. However, there will be greater attenuation in the O-band transmission. Ref. \cite{eraerds2010quantum} and Ref. \cite{patel2014quantum} performed intriguing experiments where the channel interval between the quantum and classical communication channels was 200 GHz and 300 GHz. Then Ref. \cite{wang2017long} realized the multiplexing and long-distance copropagation of QKD and terabit classical data transmission. Recently, Ref. \cite{mao2018integrating} presented the integration of QKD with a commercial backbone network of 3.6 Tbps classical data.

As data traffic demand in access and backbone networks has been increased exponentially \cite{yu2013transmission}, multiplexing over different degrees of freedom in conventional single-core single-mode fiber (SSMF), including wavelength, phase, time and polarization multiplexing, is being utilized to circumvent the future information capacity crunch \cite{essiambre2010capacity}. However, the capacity of existing standard SSMF may no longer satisfy the growing capacity demand and is approaching its fundamental limit around 100 Tbps owing to the limitation of amplifier bandwidth, nonlinear noise, and fiber fuse phenomenon \cite{qian2011101}. In order to further increase the fiber capacity, space-division multiplexing (SDM) has been proposed and attracted intensive research efforts as a solution to the problem of saturation of the capacity of conventional SSMF  \cite{saitoh2013multicore,richardson2013space,winzer2014making}. Several types of fibers have been proposed to realize SDM, such as weakly coupled multicore fiber (WC-MCF) \cite{hayashi2011design}, strongly coupled multicore fiber (SCMCF) \cite{arik2013coupled}, few mode fiber (FMF) \cite{kitayama2017few}, and few mode multicore fiber (FM-MCF) \cite{sakaguchi2016large}. Among these fibers, WC-MCF has the advantage of relatively low crosstalk \cite{saitoh2016multicore}, which is suitable for transmitting susceptible quantum signals.

In 2016, Ref. \cite{dynes2016quantum} presented the first QKD experiment in multicore fiber (MCF). They proposed an experiment in which a quantum signal and classical signal were transmitted together in a MCF. Negligible degradation in the performance of QKD was shown in their scheme since they set the wavelength of the quantum signal far away from that of classical signals. However, in a large capacity transmission system, this transmission scheme is obviously a waste of wavelength resources.

To this end, we propose a wavelength-space division multiplexing (WSDM) scheme of QKD and classical signals. Quantum signals are transmitted in one outer core while other cores are used to transmit classical signals. Dense wavelength division multiplexing (DWDM) is used in each core and separation between quantum channels and the nearest classical channels can reach 50 GHz.  The main impairment of QKD in our scheme is the intercore crosstalk (IC-XT) from the classical signals. In order to alleviate IC-XT, we propose the quantum-classical interleave scheme where the wavelengths of quantum signals and those of classical signals are interleaved. The power of IC-XT is measured without quantum signals to achieve the properties of IC-XT. The results show that the total power of classical signals in the nearest cores mainly determines the power of IC-XT in quantum core. Also, counter-propagation is analyzed to be a better co-existence method than co-propagation due to its relatively low IC-XT. Transmission distance for more than 40 km was proved to be feasible based on the measurement results. Finally, we perform experiments on a QKD system. The system can run stably for a long time in the presence of two classical signals which are 100 GHz separated from the quantum signal at 1549.32 nm. The experiment results prove that QKD is almost immune to IC-XT for counter-propagation, with the quantum bit error rate (QBER) of lower than 1\%. Also, the experiment results validate the feasibility for the transmission over the range of metropolitan area networks (typically 30 km) with our scheme.

\section{Quantum-classical interleave scheme}
\begin{figure}[h]
\centering\includegraphics[width=8cm]{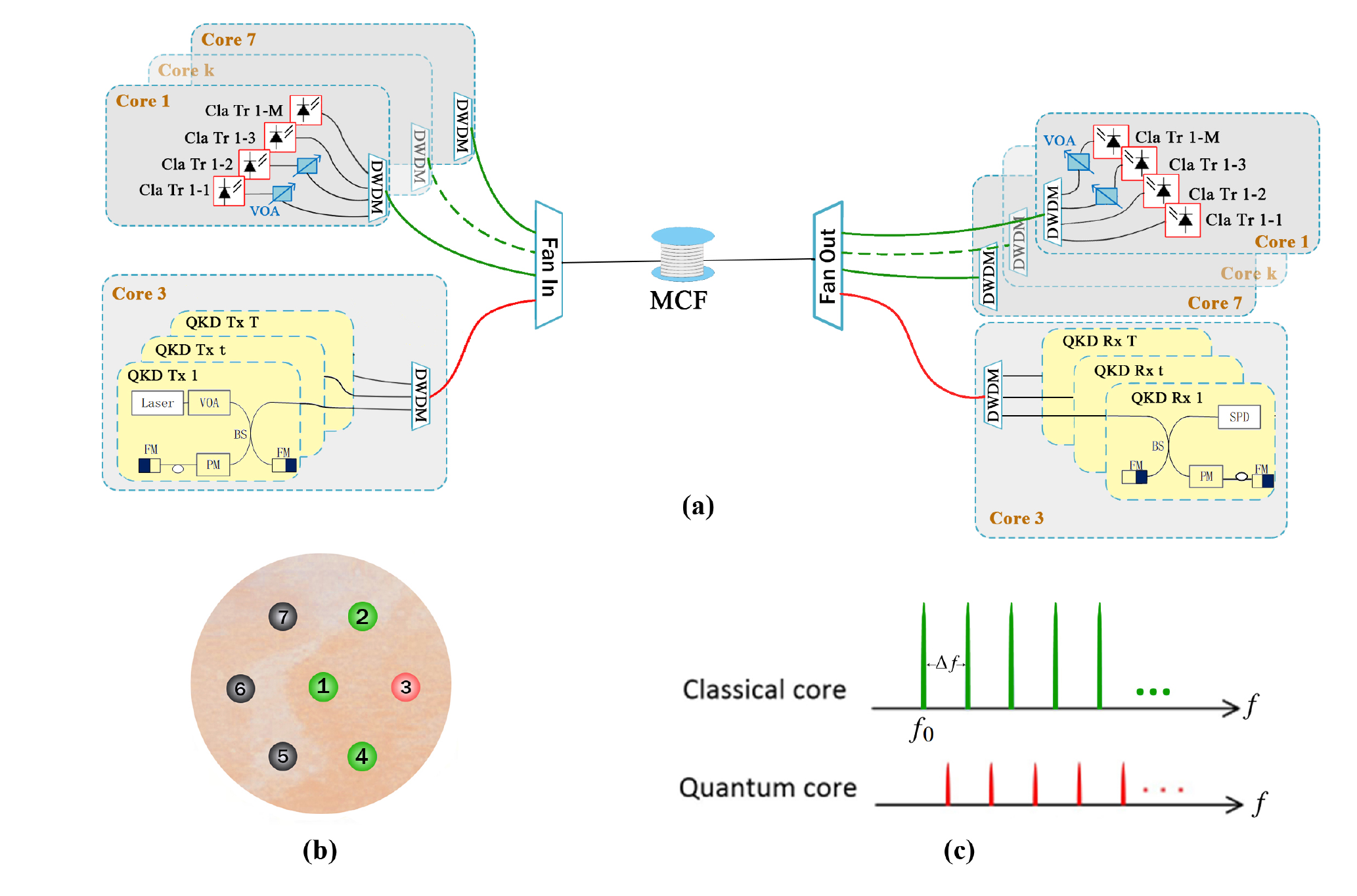}
\caption
{\label{fig:architecture}(a) MCF-based WSDM QKD system architecture. Cla Tr represents the classical tranceiver. (b) Cross-section of the seven-core fiber. (c) Quantum-classical interleave scheme.}
\end{figure}

Fig.~\ref{fig:architecture}(a) shows the MCF-based QKD system architecture supporting multiple QKD users. Homogeneous weakly coupled seven-core fiber is used in the system. As shown in Fig.~\ref{fig:architecture}(b), the inner core is set in the central of the fiber with the hexagonal disposal of the six outer cores. 

First of all, we need to allocate cores reasonably. Since the nonlinear noise from classical signals, such as Raman scattering and four-wave-mixing effect, has a stronger impact on quantum signals in the same core, quantum signals and classical signals are placed in different cores. Considering the impacts from the nearest neighbour cores are much larger than those from non-nearest neighbour, we choose one outer core to transmit quantum signals, such as core-3 in Fig.~\ref{fig:architecture}(b). This is because outer cores have three nearest neighbour cores, which is much less than the center core with six nearest neighbour cores.

When qunatum signals are placed in a designated core, the main impairment to QKD is the IC-XT from classical cores. IC-XT is the power coupling between different cores and its power is mainly concentrate on the spectral peak around the frequency of the source signal. Thus the frequencies in occupation of classical signals can not be used for quantum signals. In order to alleviate the effect of IC-XT on quantum signals, we propose the quantum-classical interleave scheme. For classical signals, we assume the first frequency of the available channel is $f_{c}(1)=f_{0}$ and the frequency spacing of two adjacent channels is $\Delta f$ in the DWDM system. So the frequency of the $n-th (n=1,2,\cdots)$ classical channel is 
\begin{equation}
\label{eq:clafre}
f_{c}(n) = f_{0}+(n-1)*\Delta f.
\end{equation}
Then we set the frequency of the $k-th (k=1,2,\cdots)$ quantum channel to
\begin{equation}
\label{eq:clafre}
f_{q}(k) = f_{0}+(k-\frac{1}{2} \;)*\Delta f,
\end{equation}
as shown in Fig.~\ref{fig:architecture}(c). In this wavelength assignment scheme, IC-XT is turned into out-of-band noise, which can be reduced by filters.

\section{Impairment Source Analysis}
\label{section:analysis}

\begin{figure}[h!]
\centering\includegraphics[width=9cm]{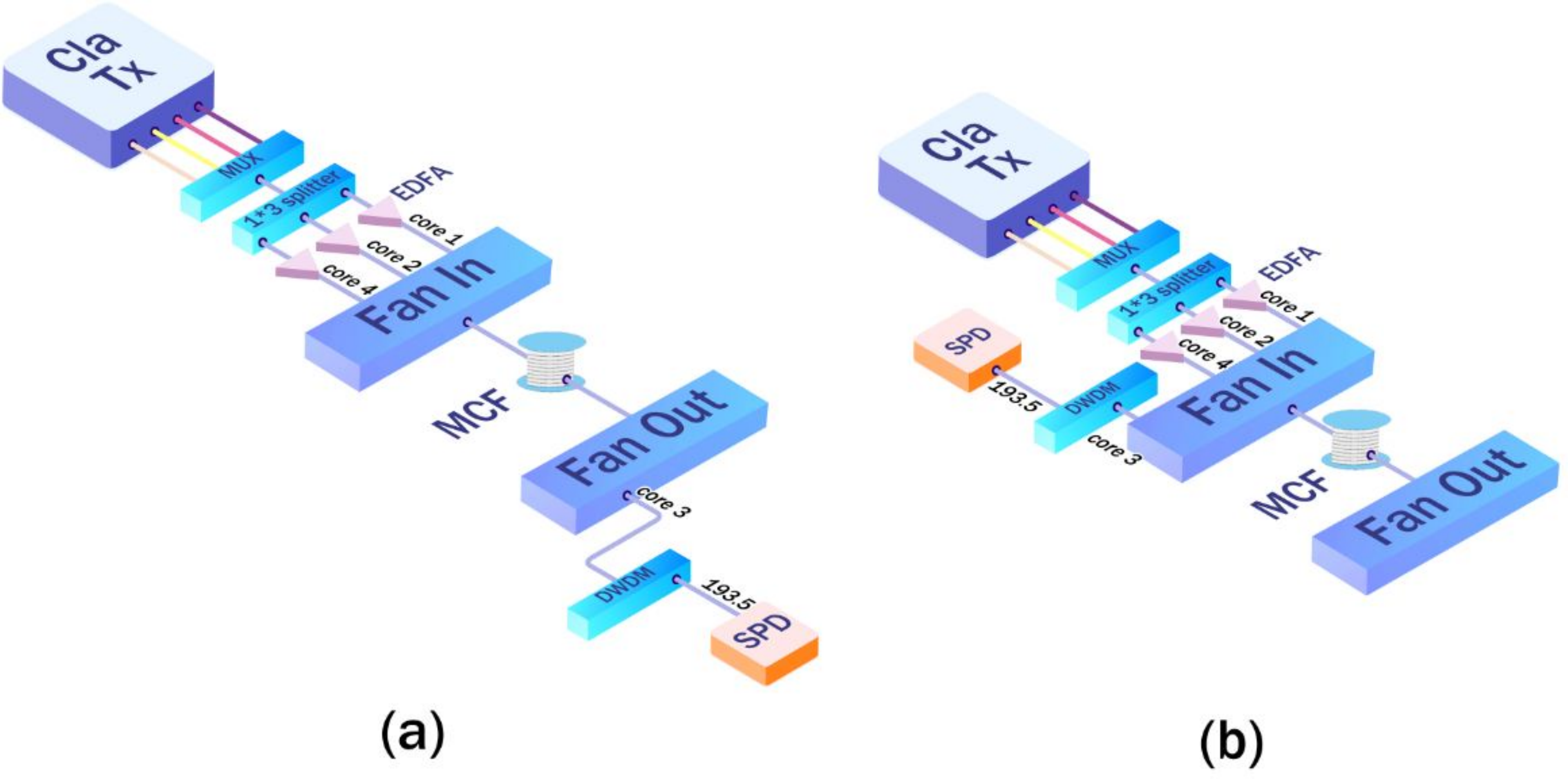}
\caption
{\label{fig:analysis}(a) Co-propagation. (b) Counter-propagation. By sending classical signals through the MFC along either the same direction of QKD signals (co-propagation) or the opposite direction (counter-propagation), photons are recorded by the SPD. However, no QKD signal is transmitted during this process in order to measure the noise.}
\end{figure}

The main impairment of the system is the IC-XT. The power of the IC-XT is the decisive factor for the QBER and secure key rate (SKR) of QKD. In this section, we measure the power of IC-XT and analyze the performance of the scheme based on the measurement results. Fig.~\ref{fig:analysis} shows our experimental setup. In Fig.~\ref{fig:analysis}(a), four classical signals are sent by the classical transmitter (Cla Tx) and enter core-1, -2, and -4 after getting through the splitter and erbium-doped fiber amplifier (EDFA). The 1-km, 7-core MCF spool forms the optical channel. A 7*1 Fan-In and a 1*7 Fan-Out are used to couple optical signals into cores at each end of the MCF. A DWDM module is connected to core-3 followed by a single photon detector (SPD). The SPD will be connected to different ports of the DWDM to measure noise at different frequencies. Fig.~\ref{fig:analysis}(b) shows the setup to measure the noise for counter-propagation and the SPD is used to detect noise at the input port of Fan-in.

\begin{figure}[h!]
\centering\includegraphics[width=9cm]{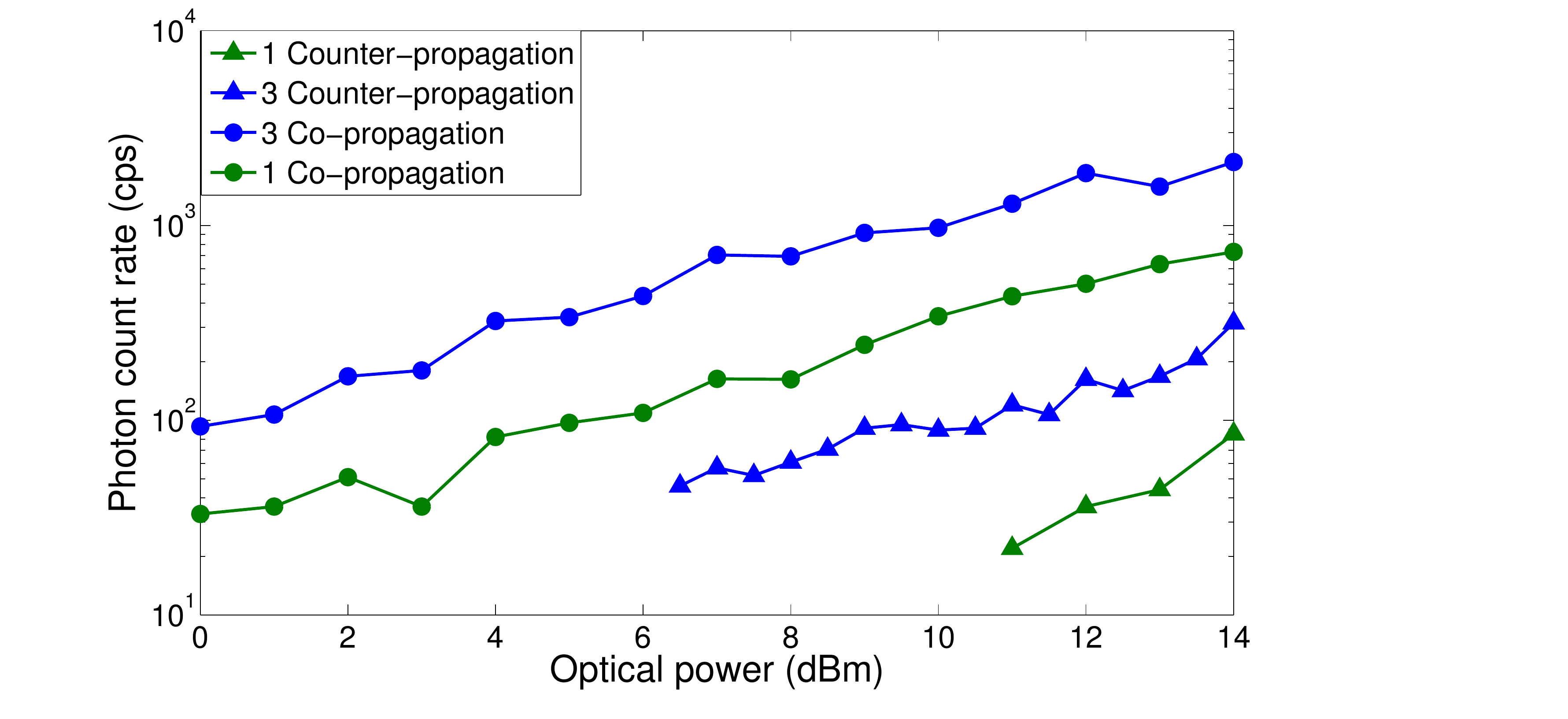}
\caption
{\label{fig:1and3}The blue lines with solid circles or triangles show the PCRs of co-propagation or counter-propagation, respectively, when classical signals are injected into core-1, -2, and -4 simultaneously. The green lines with solid circles or triangles represent the PCRs of co-propagation or counter-propagation when classical signals are only injected into core-1.}
\end{figure}
Firstly, four classical signals are sent by the classical transmitter at frequencies of 193.35, 193.45, 193.55 and 193.65 THz. After amplified by the three EDFAs, the signals are injected into core-1, -2 and -4 simultaneously or just injected into core-1. The center wavelength of each DWDM channel is on the International Telecommunication Union (ITU) grid spaced by 100 GHz. The passband of each channel is 0.22 nm with the insertion loss of 2.9 dB. Isolation of non-adjacent channels is above 50 dB, while that for adjacent channels is relatively worse, about 40 dB. SPD is connected to channel-35 of DWDM and detects photons at the frequency of 193.5 THz realized by the filtering properties of the DWDM module. The SPD operates at 20 MHz with a detection efficiency of 10\% and a dark count rate of 10 Hz in average. The detector effective gating width is set to 2.1 ns. Results are shown in Fig.~\ref{fig:1and3} in which the optical power refers to the power of classical signals at each input port of the Fan-in. Each point represents the average value of half an hour interval, which is the same with other figures in Sec.~\ref{section:analysis}. As can be seen, the photon count rates (PCRs) of both counter-propagation and co-propagation increase linearly with the input power, which means the power of the classical signals is a vital factor affecting the system. Also, the PCR of counter-propagation is about 10 dB lower than that of co-propagation. This makes counter-propagation a better way for quantum signals to co-exist with classical signals. Under the same optical power, the noise measured in the case that the three nearest cores transmit classical signals at the same time is about three times that measured in the case only one core is used to transmit classical signals. This is because the IC-XT from each classical core to the quantum core is approximately the same. In other words, the total IC-XT experienced by the quantum core is the sum of the IC-XT from the nearest neighbour cores. It can experimentally verify that the quantum signals should be placed in the outer core.

\begin{figure}[h!]
\centering\includegraphics[width=9.5cm]{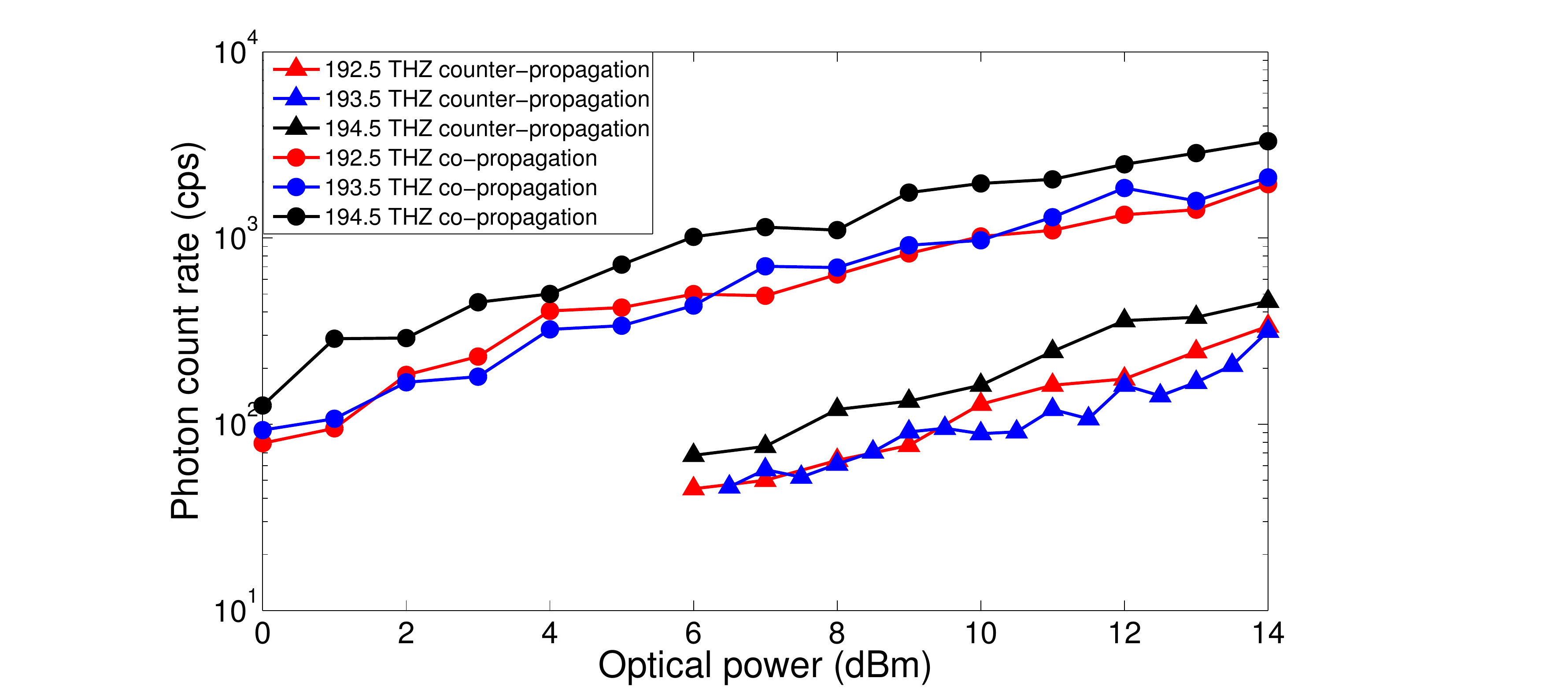}
\caption
{\label{fig:wavelength}Classical signals are injected into core-1, -2, and -4 simultaneously in these measurement. For the red line, the frequencies of four classical signals are 192.35, 192.45, 192.55 and 192.65 THz, respectively. The SPD is connected to the channel-25 of the ITU DWDM grid (192.5 THz). For the black line, the frequencies of four classical signals are 194.35, 194.45, 194.55 and 194.65 THz and the SPD is connected to the channel-45 (194.5 THz).}
\end{figure}

We also consider the case when the quantum signal is placed at other frequencies, such as 192.5 and 194.5 THz. As can be seen in Fig.~\ref{fig:wavelength}, the PCR is also dependent on optical power. However, the IC-XT is roughly the same at different wavelengths since the values of PCR at 192.5 THz are approximately equal to those at 193.5 THz for both co-propagation and counter-propagation. The values of PCR at 194.5 THz are slightly larger than those at 192.5 THz and 193.5 THz, which can be explained to be the slight difference between devices since another DWDM module is used in the measurement of 194.5 THz. Even though, this  discrepancy can be ignored for the QKD system. In summary, wavelength is not a decisive factor for IC-XT while the total power of the classical signals determine the power of IC-XT. This can also prove that it is reasonable to replace the classical signals of multiple wavelengths with the classical signal of four wavelengths in this experiment.

\begin{figure}[h!]
\centering\includegraphics[width=9.5cm]{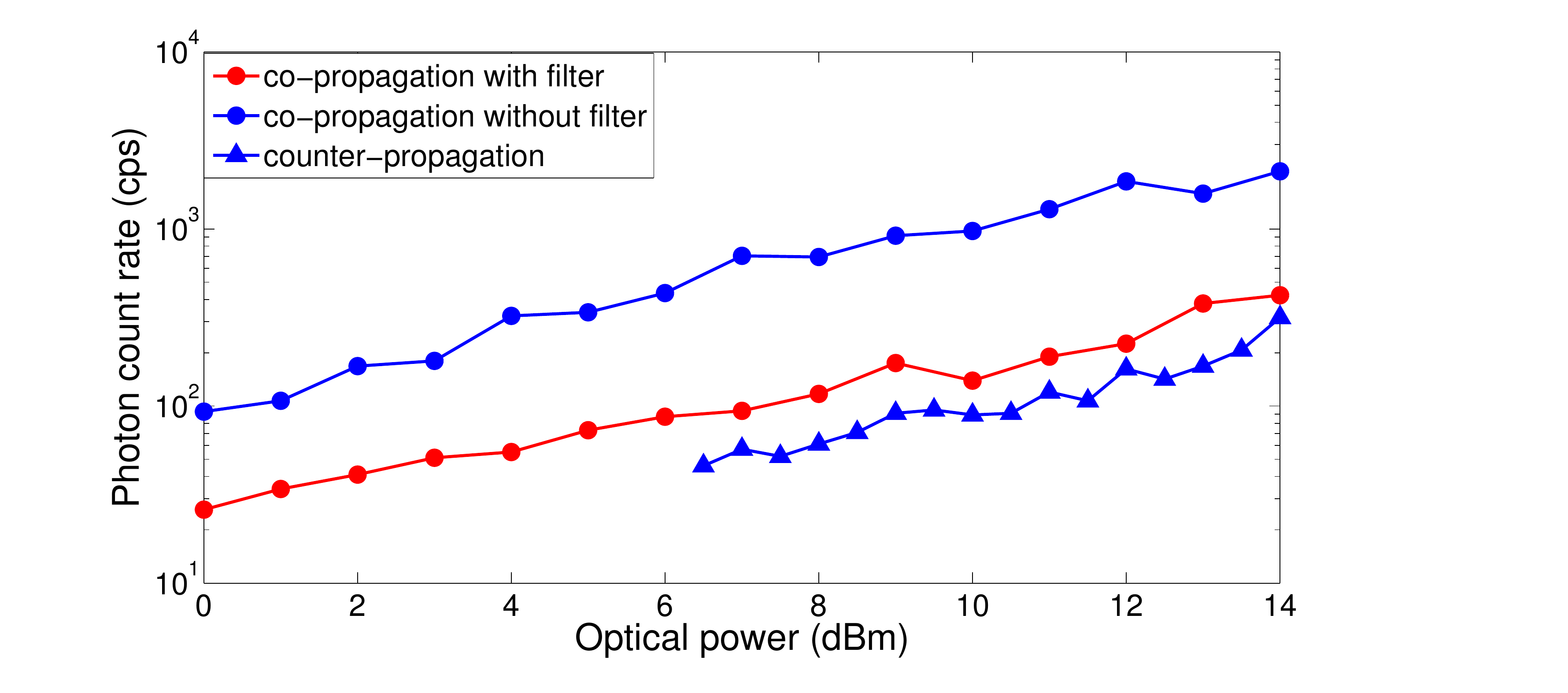}
\caption
{\label{fig:filter}The center frequency of the filter is 193.5 THz with the passband of 0.6 nm. The insertion loss of the filter is measured to be 2.1 dB. Classical signals are injected into core-1, -2, and -4 simultaneously in these measurement.}
\end{figure}

\begin{figure}[h!]
\centering\includegraphics[width=9cm]{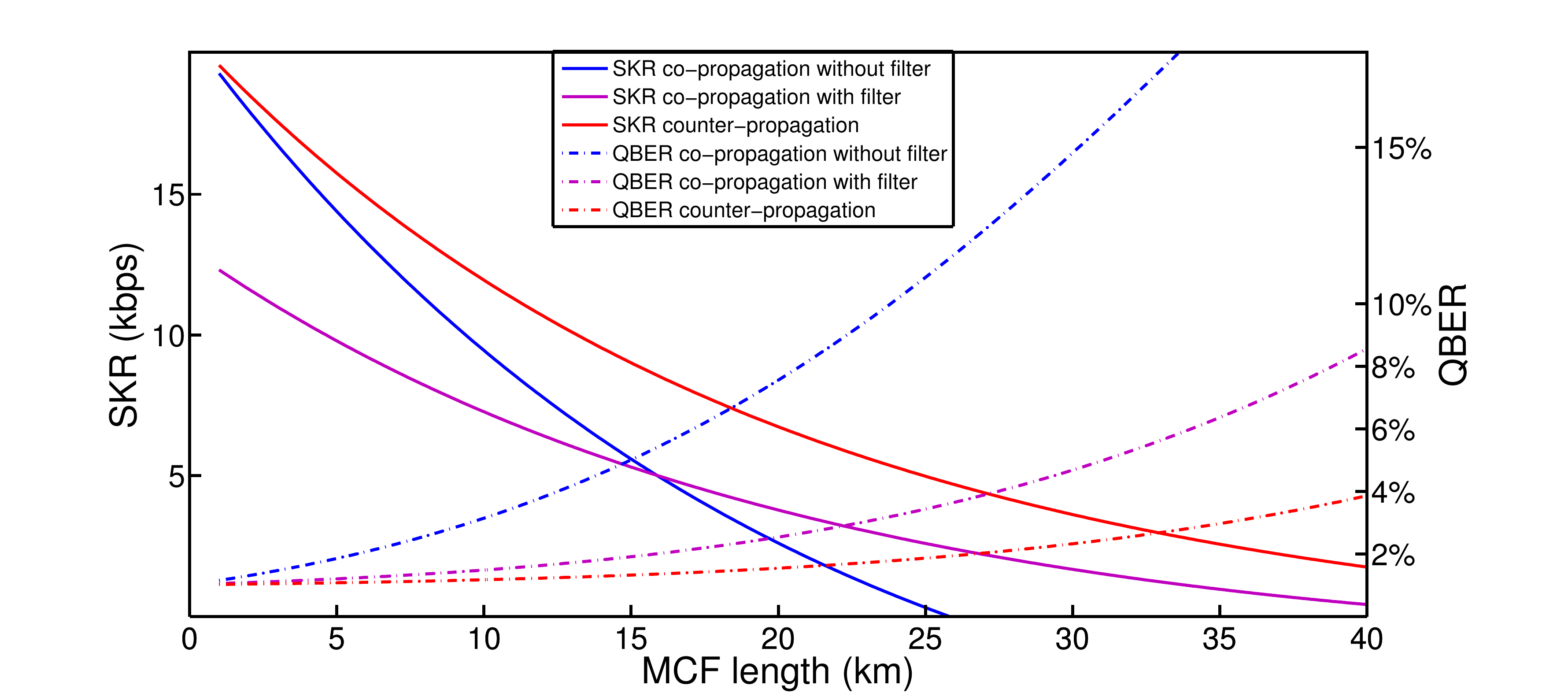}
\caption
{\label{fig:LSKRQBER}The attenuation of MCF was set to be 0.23 dB/km.}
\end{figure}

As can be seen from the above experimental results, the noise for co-propagation is relatively large. In order to alleviate the IC-XT of co-propagation, an extra filter is inserted between the SPD and the DWDM module. As shown in Fig.~\ref{fig:filter}, the noise is effectively alleviated. However, the attenuation of the quantum channel is increased due to the filter. In order to understand the impact on the performance of the system after introducing the filter, we performed the simulation shown in Fig.~\ref{fig:LSKRQBER}. The theoretical simulation was performed based on the methods of Refs. \cite{Patel2012Coexistence} and \cite{Ma2005Practical}. The weighted average of PCRs at 0 dBm is defined as
\begin{equation}
\label{eq:ave}
\overline{PCR_l}=\frac{1}{K_l}\sum_{t=1}^{K_l} \frac{PCR_{l}(t)}{10^{P_{l}(t)/10}},
\end{equation}
where $l=1, 2, 3$ corresponds to the blue, purple and red line in Fig.~\ref{fig:filter}. $P_{l}(t)$ is the optical power in the form of dBm and $PCR_{l}(t)$ is the corresponding PCR. The power of classical signals is set to 0 dBm in the simulation and we use the three weighted averages of PCR calculated by Eq.~\ref{eq:ave}. Also, the power of IC-XT is set to increase linearly with the fiber length \cite{hayashi2011design,Gan2017Investigation,Cartaxo2017Discrete}. As can be seen, the best method is the counter-propagation which can support transmission of more than 40 km. Extra filter is not required for co-propagation over short distance because SKR is higher without filter for the MCF length less than 15 km. However, the filter is necessary for long-haul transmission and the transmission distance can also reach 40 km with lower SKR compared with counter-propagation.

\section{Experiment}
\begin{figure}[h!]
\centering\includegraphics[width=9cm]{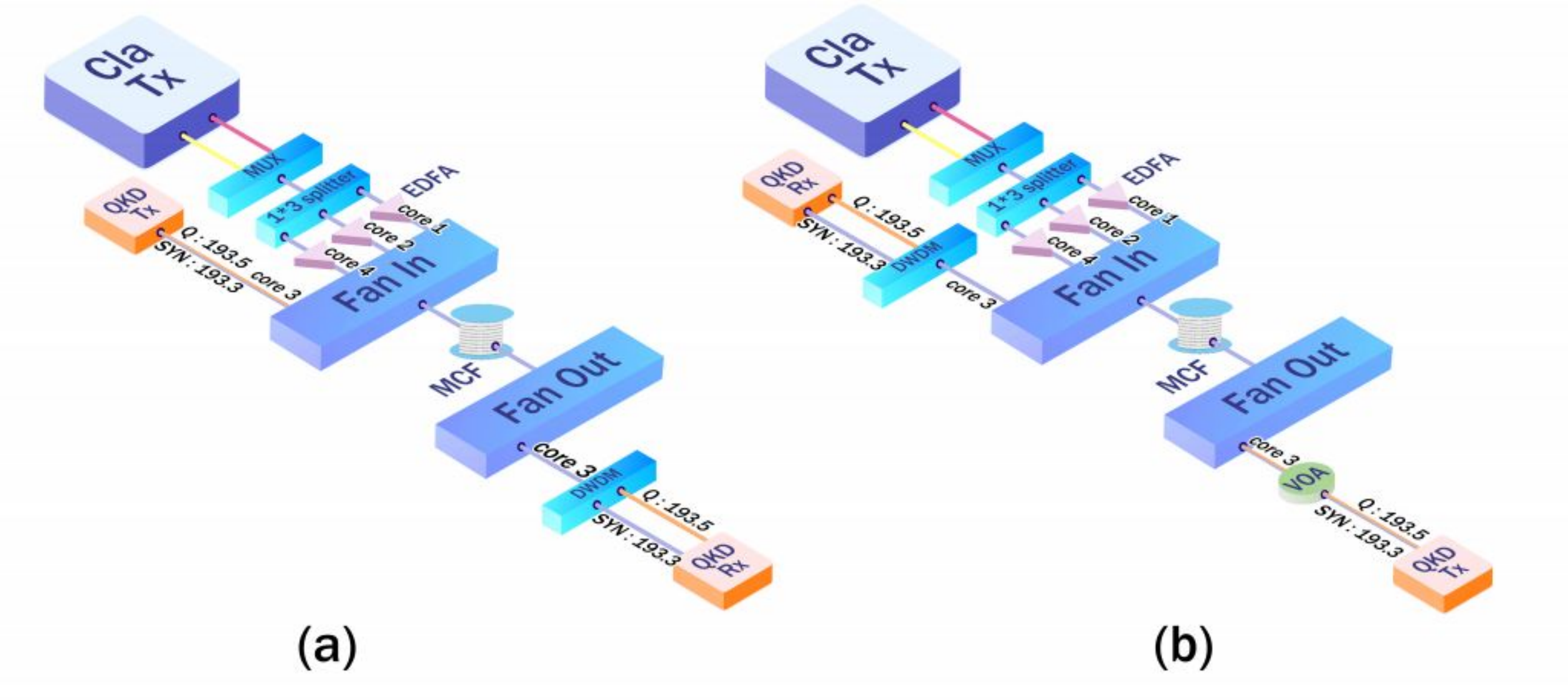}
\caption
{\label{fig:experiment}(a) Co-propagation. (b) Counter-propagation. The DWDM module is integrated in the system with the psaaband of 0.8 nm and the isolation of more than 45 dB. The variable optical attenuator (VOA) is used to simulate the attenuation of longer MCF.}
\end{figure}

We realize our scheme on a commercial QKD system. The system architecture is shown in Fig.~\ref{fig:experiment}. The quantum signal is placed at the frequency of 193.5 THz . Two classical channels are used in the system at the frequency of 193.4 THz and 193.6 THz. The synchronization signal is transmitted with quantum signal in core-3 at the frequency of 193.3 THz. The launch power of synchronization signal is adjusted to maintain a received power of about -55 dBm, which makes it harmless to quantum signal. The BB84 phase coding scheme is used in our system combined with the decoy method \cite{Wang2005Beating,Ma2005Practical}. The averaged photon numbers of the signal and decoy states are chosen to be 0.6 and 0.2, respectively. Vacuum states are also adopted as decoy states. Alice launches the three types of states at a ratio of 14:1:1, while the system operates at a frequency of 50 MHz. The entire QKD postprocessing is based on Ethernet, including the error correction with a LDPC algorithm, error verification with a cyclic redundancy check.

\begin{figure}[h!]
\centering\includegraphics[width=9.5cm]{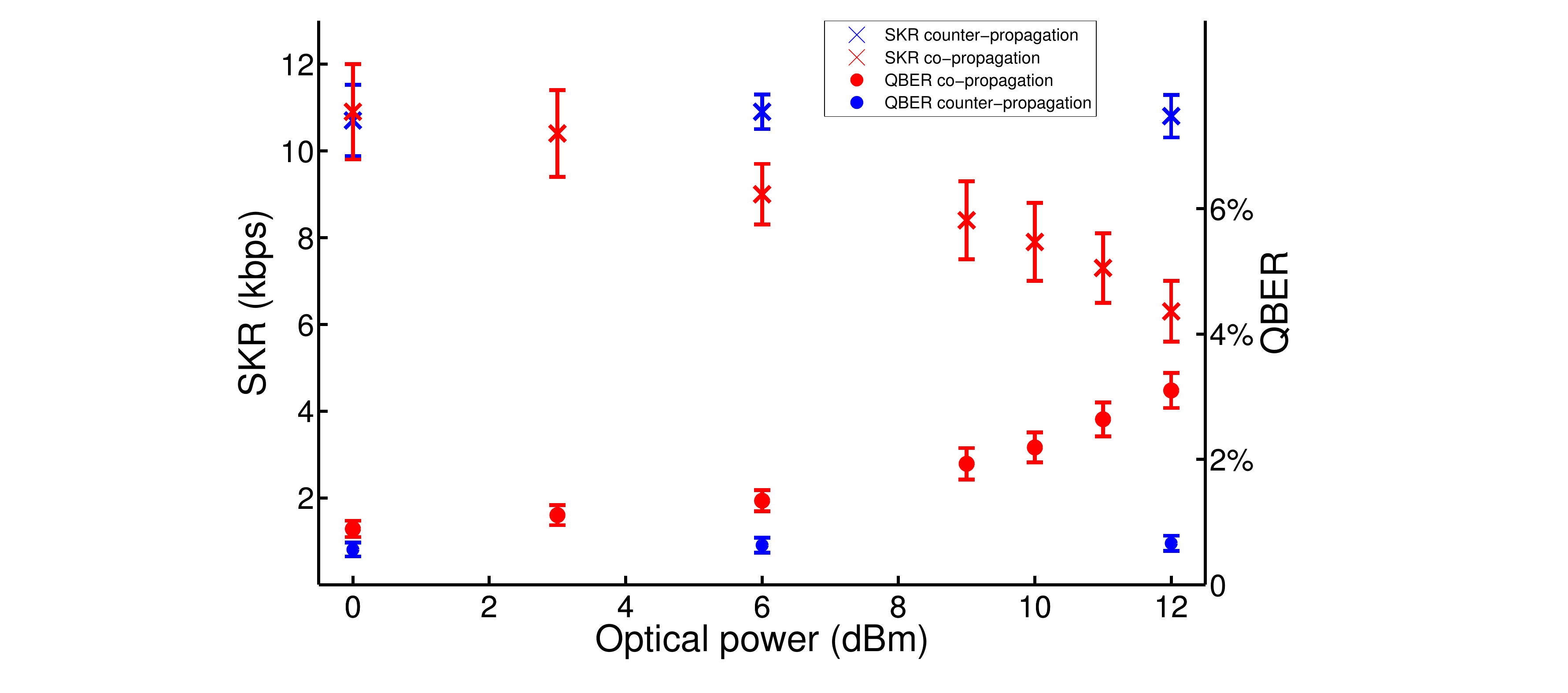}
\caption
{\label{fig:exppower}The error bars represent standard deviation at each point. Blue crosses with error bars: SKR of counter-propagation, red crosses with error bars: SKR of co-propagation, blue circles with error bars: QBER of counter-propagation, red circles with error bars: QBER of co-propagation.}
\end{figure}

Firstly, the SKR is measured to be 10.9 $\pm$ 0.55 kbps and the QBER to be 0.62 $\pm$ 0.12\% without classical signals. Then we obtain the data in Fig.~\ref{fig:exppower} by transmitting the classical signals and the quantum signal simultaneously. Every point is measured over a 24 hour period. As can be seen, the increase of optical power has little effect on the SKR and QBER of counter-propagation. Even the optical power is set to 12 dBm, the SKR hardly decrease. However, the performance of QKD will degrade when classical signals with ultra-high power are used in co-propagation. For example, the SKR will drop by 2.5 kbps with the QBER increasing by 1.3\% when the power of classical signal is 9 dBm . It is proved that counter-propagation has stronger resistance to IC-XT which is consistent with our previous analysis shown in Sec.~\ref{section:analysis}.

\begin{figure}[h!]
\centering\includegraphics[width=10.3cm]{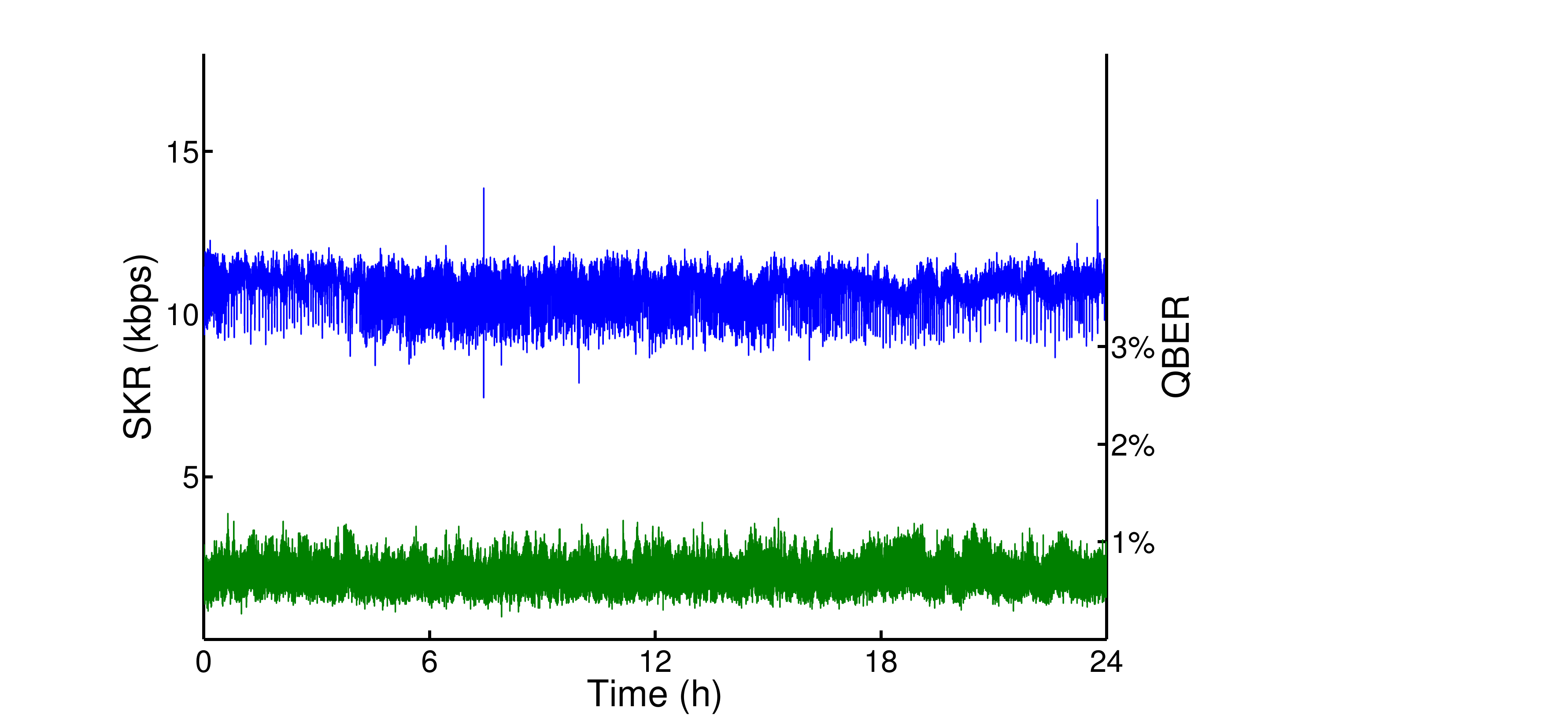}
\caption
{\label{fig:stead}The blue line plots the SKR over a 24 hour period and the green line shows the corresponding QBER.}
\end{figure}

Over the entire measurement, both the SKR and QBER display very small fluctuations around their average values with a standard deviation of less than 1.2 kbps and 0.2\%, respectively. Fig.~\ref{fig:stead} plots the QBER and SKR of counter-propagation when the optical power is 12 dBm. The average SKR is 10.8 kbps with the standard deviation of 0.49 kbps and the average QBER is 0.66\% with the standard deviation of 0.12\%. It can be seen that our scheme is a very stable solution for the WSDM QKD.

\begin{figure}[h!]
\centering\includegraphics[width=9cm]{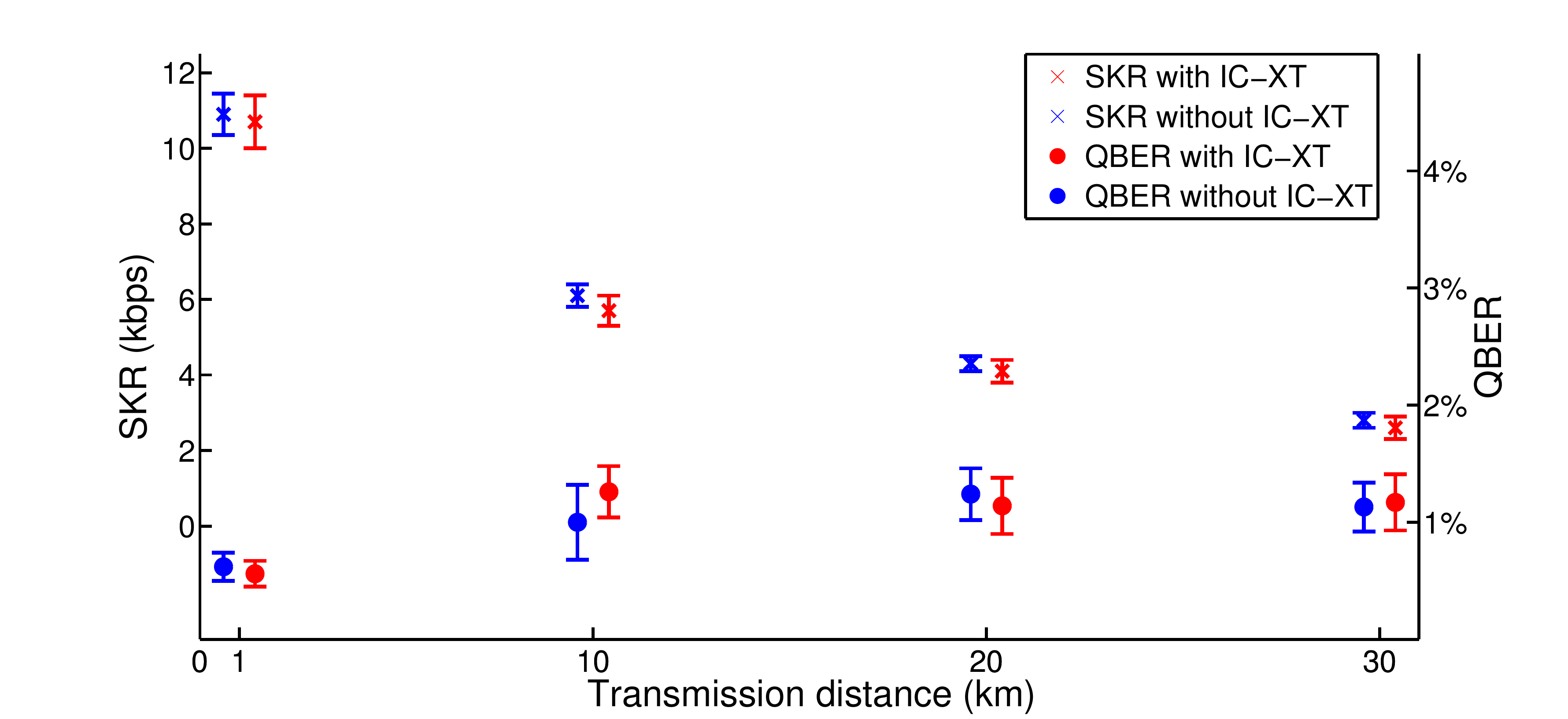}
\caption
{\label{fig:simuL}The red symbols represent the SKR and QBER of counter-propagation for 1, 10, 20, 30 km. The blue symbols show the SKR and QBER without IC-XT which means only VOA is used to simulating the attenuation and no classical signal is transmitted.}
\end{figure}

For better evaluating the performance of the scheme, it is desirable to use long fiber spools because they provide a more realistic scenario of IC-XT and attenuation. As we do not have longer MCF, the VOA and classical signals with higher power are used to simulate the experiments. The VOA is used to simulate the attenuation of MCF. Classical signals with higher power are used because the IC-XT increases linearly with the length of MCF.
We simulate the QKD counter-propagating for 10, 20, 30 km with 0 dBm classical signals. The VOA is set to 2.3, 4.6 and 6.9 dB, respectively. The power of the classical signals is correspondingly set to 10, 13, 14.8 dBm. Results in Fig.~\ref{fig:simuL} show the relatively large effect on the SKR caused by distance. A set of comparative experiments is conducted to estimate the effect of IC-XT. No classical signal is transmitted in the comparative experiments while VOA is also used to simulate the attenuation of MCF. As can be seen in Fig.~\ref{fig:simuL}, the SKR increases by 0.4 kbps at most for the same transmission distance when IC-XT is removed. It is thus clear that the decrease of SKR is mainly caused by the attenuation of the MCF which is inevitable in long-distance transmission rather than the IC-XT.

\section{Conclusion}
In this paper, we propose a WSDM scheme of quantum signals and classical signals. To alleviate the IC-XT, cores and wavelengths are allocated reasonably for quantum signals. The wavelength is assigned following the proposed quantum-classical interleave scheme where the separation between the quantum channel and the nearest classical channels can reach 50 GHz. It can save bandwith resources greatly in large capacity transmission system. Then counter-propagation is analyzed to be a better co-existence method than co-propagation based on the measurement results, which can support transmission distance of more than 40 km. Also, QKD experiments are performed in the presence of two classical channels. The experiment results prove that counter-propagation is almost immune to IC-XT and long-distance transmission is feasible. Our solution provides the possibility for QKD to be used in future large capacity transmission systems based on MCF, such as metropolitan area networks and data center.


%





\ifCLASSOPTIONcaptionsoff
  \newpage
\fi



%


\bibliographystyle{IEEEtran}
\bibliography{IEEEabrv,bare_jrnl}

\begin{thebibliography}{10}
\providecommand{\url}[1]{#1}
\csname url@samestyle\endcsname
\providecommand{\newblock}{\relax}
\providecommand{\bibinfo}[2]{#2}
\providecommand{\BIBentrySTDinterwordspacing}{\spaceskip=0pt\relax}
\providecommand{\BIBentryALTinterwordstretchfactor}{4}
\providecommand{\BIBentryALTinterwordspacing}{\spaceskip=\fontdimen2\font plus
\BIBentryALTinterwordstretchfactor\fontdimen3\font minus
  \fontdimen4\font\relax}
\providecommand{\BIBforeignlanguage}[2]{{%
\expandafter\ifx\csname l@#1\endcsname\relax
\typeout{** WARNING: IEEEtran.bst: No hyphenation pattern has been}%
\typeout{** loaded for the language `#1'. Using the pattern for}%
\typeout{** the default language instead.}%
\else
\language=\csname l@#1\endcsname
\fi
#2}}
\providecommand{\BIBdecl}{\relax}
\BIBdecl

\bibitem{Bennet1984Quantum}
C.~H. Bennet, ``Quantum cryptography : Public key distribution and coin
  tossing,'' in \emph{Proc. Of IEEE International Conference on Computers,
  Systems, and Signal processing}, 1984, pp. 175--179.

\bibitem{Gisin2001Quantum}
N.~Gisin, G.~Ribordy, W.~Tittel, and H.~Zbinden, ``Quantum cryptography,''
  \emph{Rev.mod.phys}, vol.~74, no.~1, pp. 145--195, 2001.

\bibitem{Shor2000Simple}
P.~W. Shor and J.~Preskill, ``Simple proof of security of the bb84 quantum key
  distribution protocol,'' \emph{Physical Review Letters}, vol.~85, no.~2, pp.
  441--444, 2000.

\bibitem{575910}
P.~D. Townsend, ``Simultaneous quantum cryptographic key distribution and
  conventional data transmission over installed fibre using wavelength-division
  multiplexing,'' \emph{Electronics Letters}, vol.~33, no.~3, pp. 188--190, Jan
  1997.

\bibitem{nweke2005experimental}
N.~Nweke, P.~Toliver, R.~Runser, S.~McNown, J.~Khurgin, T.~Chapuran,
  M.~Goodman, R.~Hughes, C.~Peterson, K.~McCabe \emph{et~al.}, ``Experimental
  characterization of the separation between wavelength-multiplexed quantum and
  classical communication channels,'' \emph{Applied Physics Letters}, vol.~87,
  no.~17, p. 174103, 2005.

\bibitem{chapuran2009optical}
T.~Chapuran, P.~Toliver, N.~Peters, J.~Jackel, M.~Goodman, R.~Runser,
  S.~McNown, N.~Dallmann, R.~Hughes, K.~McCabe \emph{et~al.}, ``Optical
  networking for quantum key distribution and quantum communications,''
  \emph{New Journal of Physics}, vol.~11, no.~10, p. 105001, 2009.

\bibitem{choi2011quantum}
I.~Choi, R.~J. Young, and P.~D. Townsend, ``Quantum information to the home,''
  \emph{New Journal of Physics}, vol.~13, no.~6, p. 063039, 2011.

\bibitem{eraerds2010quantum}
P.~Eraerds, N.~Walenta, M.~Legr{\'e}, N.~Gisin, and H.~Zbinden, ``Quantum key
  distribution and 1 gbps data encryption over a single fibre,'' \emph{New
  Journal of Physics}, vol.~12, no.~6, p. 063027, 2010.

\bibitem{patel2014quantum}
K.~Patel, J.~Dynes, M.~Lucamarini, I.~Choi, A.~Sharpe, Z.~Yuan, R.~Penty, and
  A.~Shields, ``Quantum key distribution for 10 gb/s dense wavelength division
  multiplexing networks,'' \emph{Applied Physics Letters}, vol. 104, no.~5, p.
  051123, 2014.

\bibitem{wang2017long}
L.-J. Wang, K.-H. Zou, W.~Sun, Y.~Mao, Y.-X. Zhu, H.-L. Yin, Q.~Chen, Y.~Zhao,
  F.~Zhang, T.-Y. Chen \emph{et~al.}, ``Long-distance copropagation of quantum
  key distribution and terabit classical optical data channels,''
  \emph{Physical Review A}, vol.~95, no.~1, p. 012301, 2017.

\bibitem{mao2018integrating}
Y.~Mao, B.-X. Wang, C.~Zhao, G.~Wang, R.~Wang, H.~Wang, F.~Zhou, J.~Nie,
  Q.~Chen, Y.~Zhao \emph{et~al.}, ``Integrating quantum key distribution with
  classical communications in backbone fiber network,'' \emph{Optics express},
  vol.~26, no.~5, pp. 6010--6020, 2018.

\bibitem{yu2013transmission}
J.~Yu, Z.~Dong, H.-C. Chien, Z.~Jia, X.~Li, D.~Huo, M.~Gunkel, P.~Wagner,
  H.~Mayer, and A.~Schippel, ``Transmission of 200 g pdm-csrz-qpsk and pdm-16
  qam with a se of 4 b/s/hz,'' \emph{Journal of Lightwave Technology}, vol.~31,
  no.~4, pp. 515--522, 2013.

\bibitem{essiambre2010capacity}
R.-J. Essiambre, G.~Kramer, P.~J. Winzer, G.~J. Foschini, and B.~Goebel,
  ``Capacity limits of optical fiber networks,'' \emph{Journal of Lightwave
  Technology}, vol.~28, no.~4, pp. 662--701, 2010.

\bibitem{qian2011101}
D.~Qian, M.-F. Huang, E.~Ip, Y.-K. Huang, Y.~Shao, J.~Hu, and T.~Wang,
  ``101.7-tb/s (370$\times$ 294-gb/s) pdm-128qam-ofdm transmission over
  3$\times$ 55-km ssmf using pilot-based phase noise mitigation,'' in
  \emph{National Fiber Optic Engineers Conference}.\hskip 1em plus 0.5em minus
  0.4em\relax Optical Society of America, 2011, p. PDPB5.

\bibitem{saitoh2013multicore}
K.~Saitoh and S.~Matsuo, ``Multicore fibers for large capacity transmission,''
  \emph{Nanophotonics}, vol.~2, no. 5-6, pp. 441--454, 2013.

\bibitem{richardson2013space}
D.~Richardson, J.~Fini, and L.~Nelson, ``Space-division multiplexing in optical
  fibres,'' \emph{Nature Photonics}, vol.~7, no.~5, p. 354, 2013.

\bibitem{winzer2014making}
P.~J. Winzer, ``Making spatial multiplexing a reality,'' \emph{Nature
  Photonics}, vol.~8, no.~5, p. 345, 2014.

\bibitem{hayashi2011design}
T.~Hayashi, T.~Taru, O.~Shimakawa, T.~Sasaki, and E.~Sasaoka, ``Design and
  fabrication of ultra-low crosstalk and low-loss multi-core fiber,''
  \emph{Optics express}, vol.~19, no.~17, pp. 16\,576--16\,592, 2011.

\bibitem{arik2013coupled}
S.~{\"O}. Ar{\i}k and J.~M. Kahn, ``Coupled-core multi-core fibers for spatial
  multiplexing,'' \emph{IEEE Photonics Technology Letters}, vol.~25, no.~21,
  pp. 2054--2057, 2013.

\bibitem{kitayama2017few}
K.-i. Kitayama and N.-P. Diamantopoulos, ``Few-mode optical fibers: Original
  motivation and recent progress,'' \emph{IEEE Communications Magazine},
  vol.~55, no.~8, pp. 163--169, 2017.

\bibitem{sakaguchi2016large}
J.~Sakaguchi, W.~Klaus, J.~M.~D. Mendinueta, B.~J. Puttnam, R.~S. Lu{\'\i}s,
  Y.~Awaji, N.~Wada, T.~Hayashi, T.~Nakanishi, T.~Watanabe \emph{et~al.},
  ``Large spatial channel (36-core$\times$ 3 mode) heterogeneous few-mode
  multicore fiber,'' \emph{Journal of lightwave technology}, vol.~34, no.~1,
  pp. 93--103, 2016.

\bibitem{saitoh2016multicore}
K.~Saitoh and S.~Matsuo, ``Multicore fiber technology,'' \emph{Journal of
  Lightwave Technology}, vol.~34, no.~1, pp. 55--66, 2016.

\bibitem{dynes2016quantum}
J.~Dynes, S.~Kindness, S.-B. Tam, A.~Plews, A.~Sharpe, M.~Lucamarini,
  B.~Fr{\"o}hlich, Z.~Yuan, R.~V. Penty, and A.~Shields, ``Quantum key
  distribution over multicore fiber,'' \emph{Optics express}, vol.~24, no.~8,
  pp. 8081--8087, 2016.

\bibitem{Patel2012Coexistence}
K.~A. Patel, J.~F. Dynes, I.~Choi, A.~W. Sharpe, A.~R. Dixon, Z.~L. Yuan, R.~V.
  Penty, and A.~J. Shields, ``Coexistence of high-bit-rate quantum key
  distribution and data on optical fiber,'' \emph{Physical Review X}, vol.~2,
  no.~4, pp. 773--777, 2012.

\bibitem{Ma2005Practical}
X.~Ma, B.~Qi, Y.~Zhao, and H.~K. Lo, ``Practical decoy state for quantum key
  distribution,'' \emph{Phys.rev.a}, vol.~72, no.~1, pp. 1--127, 2005.

\bibitem{Gan2017Investigation}
L.~Gan, L.~Shen, M.~Tang, C.~Xing, Y.~Li, C.~Ke, W.~Tong, B.~Li, S.~Fu, and
  D.~Liu, ``Investigation of channel model for weakly coupled multicore
  fiber,'' \emph{Optics Express}, vol.~26, no.~5, p. 5182, 2017.

\bibitem{Cartaxo2017Discrete}
A.~V.~T. Cartaxo and T.~M.~F. Alves, ``Discrete changes model of inter-core
  crosstalk of real homogeneous multi-core fibers,'' \emph{Journal of Lightwave
  Technology}, vol.~PP, no.~99, pp. 1--1, 2017.

\bibitem{Wang2005Beating}
X.~B. Wang, ``Beating the photon-number-splitting attack in practical quantum
  cryptography,'' \emph{Physical Review Letters}, vol.~94, no.~23, p. 230503,
  2005.

\end{thebibliography}

%








\end{document}